# Optical Imaging of Chemically and Geometrically Controlled Interfacial Diffusion and Redox in 2D van der Waals Space

Haneul Kang and Sunmin Ryu*

Department of Chemistry, Pohang University of Science and Technology (POSTECH), Pohang, Gyeongbuk 37673, Korea.

*Correspondence to: sunryu@postech.ac.kr

**Abstract**

Molecular motions and chemical reactions occurring in constrained space play key roles in many catalysis and energy storage applications. However, its understanding has been impeded by difficulty in detection and lack of reliable model systems. In this work, we report geometric and chemical manipulation of $O_2$ diffusion and ensuing $O_2$-mediated charge transfer (CT) that occur in the 2D space between single-layer transition metal dichalcogenides (TMDs) and dielectric substrates. As a sensitive real-time wide-field imaging signal, charge-density-dependent photoluminescence (PL) from TMDs was used. The two sequential processes inducing spatiotemporal PL change could be drastically accelerated by increasing the interfacial gap size or introducing artificial defects serving as CT reaction centers. We also show that widely varying CT kinetics of four TMDs are rate-determined by the degree of hydration required for the reactions. The reported findings will be instrumental in designing novel functional nanostructures and devices.

## Introduction

Understanding molecular behaviors in a constricted space is not only a challenging task to fundamental science but also pivotal in designing various forms of structured nanomaterials for heterogeneous catalysis,[1] energy storage,[2] drug delivery,[3] etc. While various molecular interactions with confining walls will hinder molecular motions and mass transport across the narrow space,[4] molecules may exchange charges with specific sites of the walls and undergo chemical reactions. For example, more than 100 molecular or ionic species such as metal ions, diatomic halogens, halides, chalcogenides and even benzene are known to intercalate graphite,[5] the most studied system among various layered materials. Their intercalation is driven by host-guest charge transfer (CT) and proceeds via an extremely slow diffusion as observed for alkali metals. Whereas the diffusion of the intercalants has been studied by neutron scattering,[6] nuclear magnetic resonance[7] and electrochemical methods,[8] the direct optical imaging of intercalants remains as a great challenge because of their poor sensitivity as a probe.

Spectro-microscopic contrast in the solid walls[9] can be utilized instead as such molecule-wall interactions not only transform the molecules but also impose substantial perturbation on the walls.[10] The latter effect becomes increasingly important and governs the material properties as the dimensions decrease so that the fraction of surface atoms is significant.[11] One prime example is CT chemical doping of low dimensional materials including carbon nanotubes,[12] graphene,[13] and two-dimensional (2D) transition metal dichalcogenides (TMDs)[14] used to modify their electronic,[15] optical,[16] and electrical[17] properties as done for conductive polymers.[18] Among many chemical dopants such as halogens[19] and molecules with significant electron affinity,[20] the composite electron acceptor consisting of oxygen and water ($O_2$/$H_2O$ redox couple)[21] is advantageous because of its mild reactivity and non-toxicity. Despite the early confusion and controversy over its role in the spontaneous[22-23] and activated[13] hole doping in the ambient conditions, it is established[24-25] that its action leads to the downshift of Dirac point[22] and phonon hardening[23] in graphene, and enhanced photoluminescence (PL) in TMDs[14], the third of which enabled the recent real-time observation of $O_2$ diffusion at $WS_2$-$SiO_2$ interface.[10]

The high sensitivity of the molecule-wall probe may allow a direct investigation on how the geometry and chemical environment of a constricted space affect molecular behaviors. The diffusion process will be accelerated as the constriction is relieved by increasing the gap between two planar walls.[26] One may also be able to elucidate what factors govern the CT kinetics between 2D materials and molecules. Although the molecular origin,[13] thermal activation,[24] and electrochemical nature[10, 21] of the $O_2$-driven CT have been revealed, the current understanding is insufficient to unravel the complete machinery of the phenomenon that may occur to and modify virtually any materials. In fact, the CT process is essentially open-circuit oxygen reduction reaction (ORR)[27] that consumes the electrons of 2D materials: $O_2 + 4H^+ + 4e^- \leftrightarrow 2H_2O$ (acidic), $O_2 + 2H_2O + 4e^- \leftrightarrow 4OH^-$ (basic). The thermodynamic driving for ORR heavily depends on the chemical nature of reaction centers[27] and should also be affected by the degree of hydration for systems under gaseous environments.

Herein, we report real-time imaging of $O_2$ diffusion and CT reactions that occur in 2D space between single-layer (1L) TMDs and dielectric substrates. Using in-situ wide-field PL imaging, we demonstrated that both processes could be controlled by geometric or chemical manipulation of the van der Waals (vdW) interface. By increasing the interfacial gap size or adding artificial defects in TMDs, the spatiotemporal change in PL images could be drastically accelerated because of enhanced interfacial diffusion or CT rates. This fact also revealed the pivotal role of defects as CT reaction centers. These findings will be valuable in designing novel functional nanostructures and devices.

**Methods**

***Preparation and treatment of samples***. Samples of 1L TMDs ($MX_2$) were prepared using mechanical exfoliation[22] of bulk crystals (2D semiconductors Inc.) in the ambient environment. The thickness of 1L TMDs was verified with their optical contrast and Raman spectra as shown in Fig. S7. To vary the average gap spacing between 1L TMDs and substrates, we adopted the following methods: i) for minimum-gap samples with a hydrophobic interface ($^{min}MX_{BN}$), 1L $MX_2$ was dry-transferred[28] onto thin hBN crystals (2.5 ~ 3 nm in thickness) that was supported on 285-nm $SiO_2$/Si substrates. ii) for minimum-gap samples with a hydrophilic interface ($^{min}MX_{sapp}$), 1L $MX_2$ was mechanically exfoliated onto crystalline sapphire substrates terminated with (0001) facet (MTI Corp.). iii) Intermediate-gap samples ($^{int}MX_{SiO2}$) were prepared by mechanical exfoliation onto $SiO_2$/Si substrates. iv) For maximum-gap samples ($^{max}MX_{SiO2}$), 1L $MX_2$ was dry-transferred onto $SiO_2$/Si substrates. To obtain statistical significance, we prepared more than 10 samples for each gap spacing (5 for $^{min}MX_{sapp}$) and obtained consistent results.

Before the dry transfer in i) and iv), 1L $MX_2$ was exfoliated onto PDMS (polydimethylsiloxane) substrates as described elsewhere.[28] Before sample preparation, $SiO_2$/Si and sapphire substrates were cleaned by UV-ozone (UVO) treatments (> 25 mW/cm$^2$ for 185 & 254 nm) for 5 min. Whereas the former were followed by annealing at 300 ºC for 5 minutes in the air to remove UVO-generated surface charges, the latter were similarly treated at 1000 ºC for 6 h to remove surface contamination or change the surface morphology and stoichiometry by surface reconstruction.[29-30] To generate defects in the basal-plane of 1L TMDs, samples were UVO-treated for 10 ~ 300 s.

We note that the gap spacing may vary within a given TMD sample because of its structural deformation that are caused by finite roughness of the substrates. It is known that such structural ripples have a very small correlation length of ~30 nm.[31] Considering that our laser spot size and the spatial resolution of PL imaging were larger than 500 nm, such short-ranged variation in gap spacing should be averaged out within a given focal spot.

***AFM measurements***. An atomic force microscope (Park Systems, XE-70) was utilized for topographic measurements using Si tips with a nominal tip radius of 8 nm (MicroMash, NSC-15). Tapping mode gave more reliable results than the non-contact mode that suffers from chemical contrast between different materials.[32]

***Raman and PL measurements***. Raman and PL measurements were performed with a home-built micro-spectroscopy setup described elsewhere.[33] Briefly, solid-state laser beams operated at 457.8 nm and 514.3 nm were focused onto samples within a spot size of ~1 μm using a microscope objective (40X, numerical aperture = 0.60) to obtain Raman and PL spectra, respectively. The average power for both measurements was maintained below 6 μW to avoid photoinduced damage. The backscattered signals were collected with the same objective and guided to a spectrometer combined with a liquid nitrogen-cooled CCD detector. The overall spectral accuracy was better than 1 and 5 $cm^{-1}$ for Raman and PL spectra, respectively.

For wide-field PL imaging, 514.3 nm laser was used for excitation as described elsewhere.[10] The collimated beam from the source was 3 times expanded by a Galilean beam expander (Thorlabs, BE03M-A) and focused on the back-focal plane of the above objective lens. The spatial resolution of PL images was ~500 nm (Fig. S9). As the FWHM of the illuminated area was ~220 μm, the variation of excitation intensity across 30x30 $\mu m^2$ was less than 5%. The excitation power on samples was maintained below 60 μW (~4.0x10$^{-4}$ μW/$\mu m^2$), which induced no irreversible changes.

As briefly explained in the main text, PL intensity enhancement was found to be more reliable in monitoring CT reactions than intensity itself because the latter exhibited significant spatial variation for most samples. Spatially varying charge density causing the PL intensity variation can be attributed to the inhomogeneity in interfacial hydration or defect density.

***Control of gas and liquid environments***. Spectroscopic measurements in various gas and liquid environments were performed using homemade gas and liquid-phase optical cells.[10] Ar, $O_2$ or their mixtures could be introduced at a preset flow rate into the cells: 1,000 mL/min (Ar) and 250 mL/min ($O_2$:Ar = 1:4) for the gas cell; 250 mL/min (Ar) and 250 mL/min ($O_2$) for the liquid cell. The relative humidity in the gas cell was maintained below 4% by flowing Ar for 2 h before measurements. Finite-length tubes and manifolds for gas delivery to the cells led to a response time of ~10 s. For samples immersed in the liquid cell, the net time for $O_2$ to reach the sample surface was (~30 s) longer than the response time despite vigorous sparging through water, which could be attributed to slow dissolution and diffusion in liquid.

## Results and Discussion

***Geometric and chemical manipulation of interfacial CT***. To study how the CT process is affected by the geometric dimensions and chemical nature of the vdW interface, we carefully designed samples by applying different preparation methods as described in Methods. For a minimum gap spacing, 1L TMDs of $MX_2$ (M = Mo, W; X = S, Se) was supported on crystalline substrates: hydrophobic hexagonal BN ($^{min}MX_{BN}$) and hydrophilic sapphire ($^{min}MX_{sapp}$). Intermediate and maximum spacings were achieved on amorphous $SiO_2$/Si substrates by mechanical exfoliation[22] ($^{int}MX_{SiO2}$) and dry transfer[34] ($^{max}MX_{SiO2}$) methods, respectively. The height AFM images in Fig. 1a show that the four types for $WS_2$ are highly flat except for $^{max}MX_{SiO2}$ that contains nanoscopic bubbles generated during the transfer step.[35] Height histograms from multiple samples as shown in Fig. 1b revealed that the apparent spacing defined as the difference between step height and interlayer spacing (~0.62 nm) of bulk $WS_2$ is $0.17 \pm 0.03$, $0.16 \pm 0.05$, $0.47 \pm 0.02$, and $1.64 \pm 0.35$ nm for $^{min}WS_{BN}$, $^{min}WS_{sapp}$, $^{int}WS_{SiO2}$, and $^{max}WS_{SiO2}$, respectively. Despite the small gap difference between $^{int}WS_{SiO2}$ and $^{min}WS_{sapp}$, their effective gap sizes differ more because of the atomically rough surface of $SiO_2$ substrates,[31] which is validated below. The large interfacial spacing in $^{max}MX_{SiO2}$ is attributed to air-borne molecules trapped during sample preparation.[35] It is also noted that the bare crystalline substrates have narrow distributions, whereas $^{max}WS_{SiO2}$ exhibits a broad and asymmetric one because of structural irregularities including the nanobubbles.

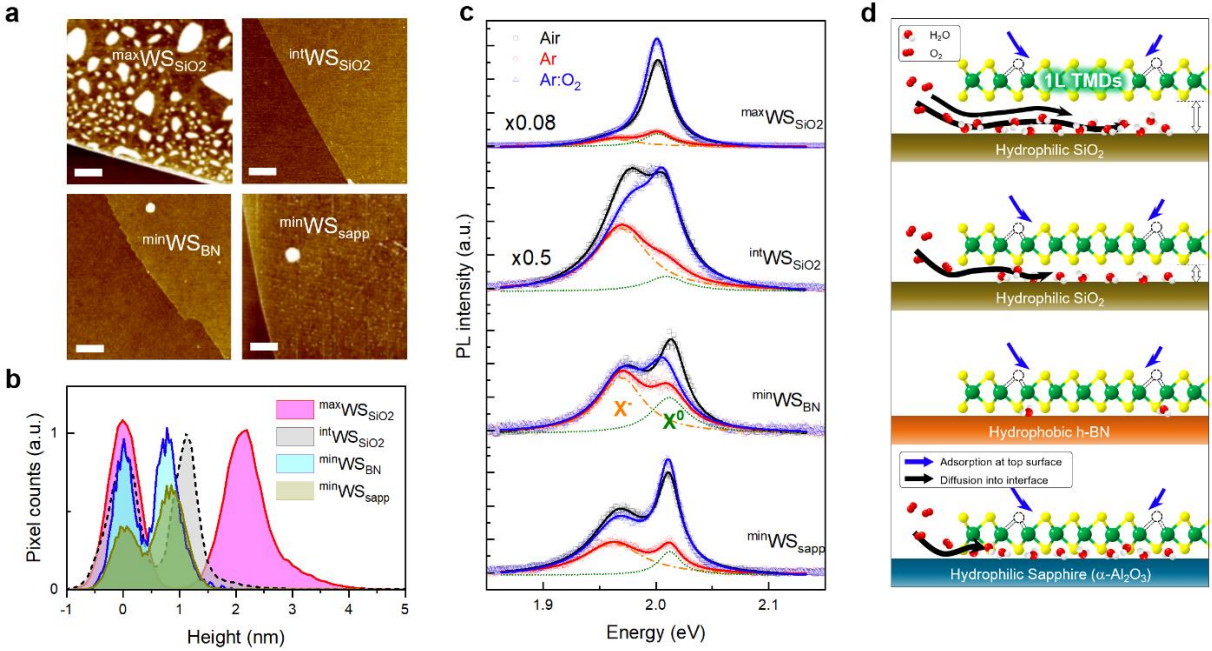

**Figure 1. Geometric and chemical control of interfacial CT reaction.** (a ~ b) AFM topographic images (a) and height histograms (b) of 1L WS₂ samples with maximum ($^{max}$WS$_{SiO2}$), intermediate ($^{int}$WS$_{SiO2}$), and minimum ($^{min}$WS$_{BN}$ & $^{min}$WS$_{sapp}$) interfacial gaps, where the scale bars are 0.5 μm. Irregular structures including the bubbles in $^{max}$WS$_{SiO2}$ were excluded in the histograms in (b). Each histogram in (b) was normalized arbitrarily to minimize visual congestion. (c) Representative PL spectra of the four-kind samples in the ambient air (black), Ar (blue), and Ar:O₂ (red) gas. Solid lines are double Lorentzian fits for excitons (X⁰) and trions (X⁻). (d) Side-view schemes for interfacial diffusion (black arrows) and surface adsorption (blue arrows) of O₂. The four sub-panels correspond to the four samples in (c) in the same order from the top.

To initiate the CT process, we introduced O₂ gas (20% in Ar) to the gas-tight optical cell containing samples after pre-equilibration with Ar gas for 2 h (see Methods).[10] The relative humidity in the cell was maintained below 4% at the moment of O₂ injection and gradually decreased during optical measurements (Fig. S1). To monitor CT and quantify the subsequent change in the charge density of TMDs, we used the ratiometric analysis of PL signals from excitons (X⁰) and trions (X$^\pm$).[36] The latter or charged excitons are formed between the former and excess charges (electrons for natively n-type WS₂) with dissociation energy ($E_{diss} = E_{X0} - E_{X-}$): X⁰ + e ↔ X⁻. Because the exoergic association reaction is governed by the mass action law,[36] the density of each species is related to an equilibrium constant, which enables the quantification of electron density (***n*₂**).[10, 37-38] As shown in Fig. 1c, the PL spectra of WS₂ consisted of two peaks, denoted X⁰ and X⁻, regardless of the gap spacing schematically shown in Fig. 1d. The intensity of the former increased when O₂ was introduced because ***n*₂** was decreased by the holes injected by the CT process. The O₂-induced change could be reversed repeatedly.[10] The decrease of E$_{diss}$ in Table S1 also supports the lowering of the Fermi level of WS₂.[41] We also note that the contribution of $X^o$ to the O₂-equilibrated spectra was the more for the larger gap spacing, which suggests an intimate relationship between the degree of CT and the interfacial dimension. In Fig. 2a, we plot the ratiometry-derived $n_e$ as a function of the PL intensity ratio ($I_{X^-}/I_{X_o}$) for Ar and O₂-equilibrated states, respectively. For all the samples, $n_e$ and $I_{X^-}/I_{X_o}$ decreased when Ar was replaced with O₂, which agreed with the spectral change induced by electrical means[39-40] (dotted and dashed lines in Fig. 2a). Interestingly, both quantities were smaller in either gas for larger gap spacing except for $^{min}WS_{sapp}$. Additionally, their attenuation was more drastic for the samples with hydrophilic substrates upon switching from Ar to O₂ as summarized in Table S1, which agrees with the nature of the CT reaction.

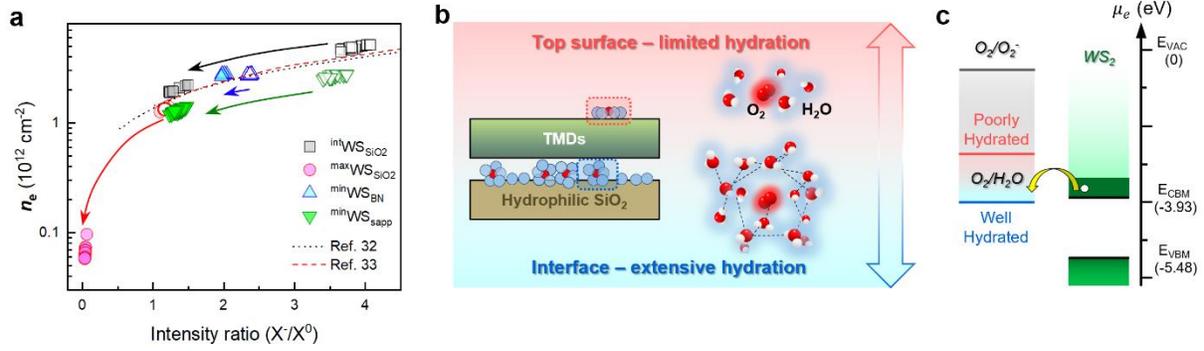

**Figure 2. Substrate-assisted CT from WS₂.** (a) Electron density ($n_e$) in Ar (open symbols) and Ar:O₂ (filled symbols) gas given as a function of PL intensity ratio ($X^-/X^0$), where the dotted lines and arrows represent electrical modulations[39-40] and the change of gas environment, respectively. Each group of symbols indicates the typical scatter observed from repeated measurements. (b) Scheme illustrating extensive and limited hydration at the hydrophilic interface and hydrophobic TMD surface, respectively. (c) Energy-level diagram for charge transfer (CT) from WS₂ to O₂/H₂O redox couples.

The gap-dependence of CT arises from the availability and energetics of the composite dopants (O₂ and H₂O) at the interface. Most of all, O₂ will be more readily available for larger gaps because the gap size of $^{min}WS_{BN}$ is only comparable to the vdW diameter of O₂ (0.38 nm).[42] As depicted in Fig. 1d, the samples with significant gap space can accommodate CT reactions at the interface as well as the top surface of WS₂. Time-lapse PL imaging verified enhanced diffusion of O₂ for larger gaps as will be shown below. Secondly, the more significant change in $n_e$ for $^{min}WS_{sapp}$ than $^{min}WS_{BN}$ despite their similar gap spacings can be attributed to the higher hydrophilicity of sapphire. Whereas dry gases of Ar and O₂ were used, the hydrophilic interfacial space in Fig. 2b may carry a significant amount of ambient water molecules,[43] unlike the top surface of pristine WS$_2$. More specifically, the CT rate can be affected by the hydration level of the reaction centers that is governed by the geometric dimension and chemical nature of the interfacial space, as depicted in Fig. 2c. Within the Marcus-Gerischer theory,[44] electrons are transferred from the donor (WS$_2$) to the acceptor (composite dopants) levels, which was supported by pH-dependent CT rates of WS$_2$ and MoS$_2$ in solutions.[10] For the current samples in dry gas, however, the degree of hydration at CT centers is much lower than in an aqueous environment, which leads to destabilization of the molecular acceptor state (Fig. 2b & 2c). In such a circumstance, the CT rate can be determined by the hydration level. We note that the gap spacing of the samples is on the order of the typical size (~0.4 nm) of the first hydration shell for small ions.[45] Then, the CT rate will be higher for larger gaps and hydrophilic substrates (thus substrate-assisted) that allow more extensively hydrated clusters.

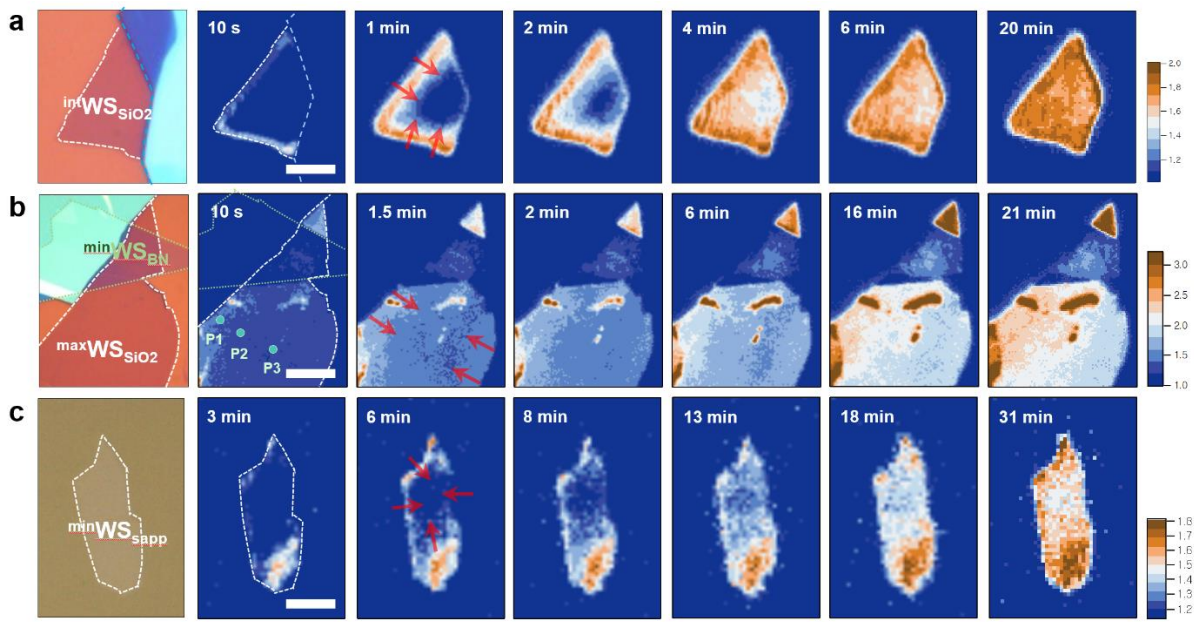

**Figure 3. Real-time imaging of gap-dependent interfacial diffusion of dopants.** (a ~ c) Optical micrographs (first column) and time-lapse wide-field PL enhancement images: $^{int}WS_{SiO2}$ (a), $^{max}WS_{SiO2}$ and $^{min}WS_{BN}$ (b), and $^{min}WS_{sapp}$ (c). The red arrows in the third images of (a ~ c) denote the direction of changes. The raw data for (a) was from a previous work.[10] The samples were pre-equilibrated with Ar gas for 2 h before exposure to Ar:O$_2$ mixed gas at time zero. P1 ~ P3 mark where the data for Fig. 5a were obtained. Scale bars are 6, 10, and 4 μm, respectively.

***Real-time wide-field PL imaging of gap-dependent CT kinetics.*** Using the high $n_e$-sensitivity of PL signals, we mapped out spatiotemporal change of CT reactions for the three samples. Figure 3 shows their PL enhancement images obtained for various $O_2$-exposure times (t), where the enhancement was defined as $(I_t - I_0)/ I_0$, where $I_0$ corresponds to the PL intensity at t = 0. As reported recently,[10] the enhancement images were more reliable and sensitive to changes than their intensity counterparts because of the significant spatial inhomogeneity of the latter as can be seen in Fig. S2. In Fig. 3a, the two edges of $^{int}WS_{SiO2}$ started to show PL enhancement within 10 s upon exposure to $O_2$, which was followed by a directional enhancement of ~1.9 for the next several min. The edge-to-center progression was distinctive of CT driven by the interfacial diffusion of the dopants.[10] For $^{max}WS_{SiO2}$, the initial progression of enhancement (~1.5) was substantially accelerated and completed within 2 minutes (Fig. 3b). An additional spatial propagation of enhancement was observed between 2 and 6 min, with the entire area finally giving 2 ~ 4 times stronger PL at 21 min. The two sets of data corroborate that faster and more intensive CT occurs for the larger gaps, which agrees with Fig. 1d.

Reduced gap spacing led to a smaller enhancement (~1.4) in $^{min}WS_{BN}$ that corresponds to the $WS_2$ area on hBN in Fig. 3b. Its non-directional progression was attributed to CT localized at the exposed surface of $WS_2$, not the interface between $WS_2$ and hBN.[10] This assignment was also validated by the amplification of the front-surface CT by increased defects, as will be shown later. Despite its reduced gap, however, $^{min}WS_{sapp}$ in Fig. 3c also exhibited edge-to-center propagation that took a few times longer than $^{int}WS_{SiO2}$. This fact indicates that the interfacial CT is governed primarily by the gap dimension and promoted for hydrophilic substrates. Sapphire (0001) facet undergoes dissociative hydroxylation at Al-rich sites with ambient water[46] and is more hydrophilic (water contact angle = ~5º)[47] than hBN (water contact angle = ~60º).[48] AFM topography indeed revealed trapped water layers and nanoscale islands[46] aligned along with steps between (0001) terraces (Fig. S3). The directional propagation (Fig. 3c) and the presence of the interfacial water supports that the latter serves as a hydration medium for the chemical species involved in the CT process. Because of the atomic flatness of (0001) terrace and its intimate contact with TMDs,[46] the 1D channel between $WS_2$ and the step edges may provide a route for $O_2$ diffusion as suggested for water.[46] Remarkably, the diffusion-mediated CT was observed for all of the four TMDs (Fig. S4), which suggests its universality.

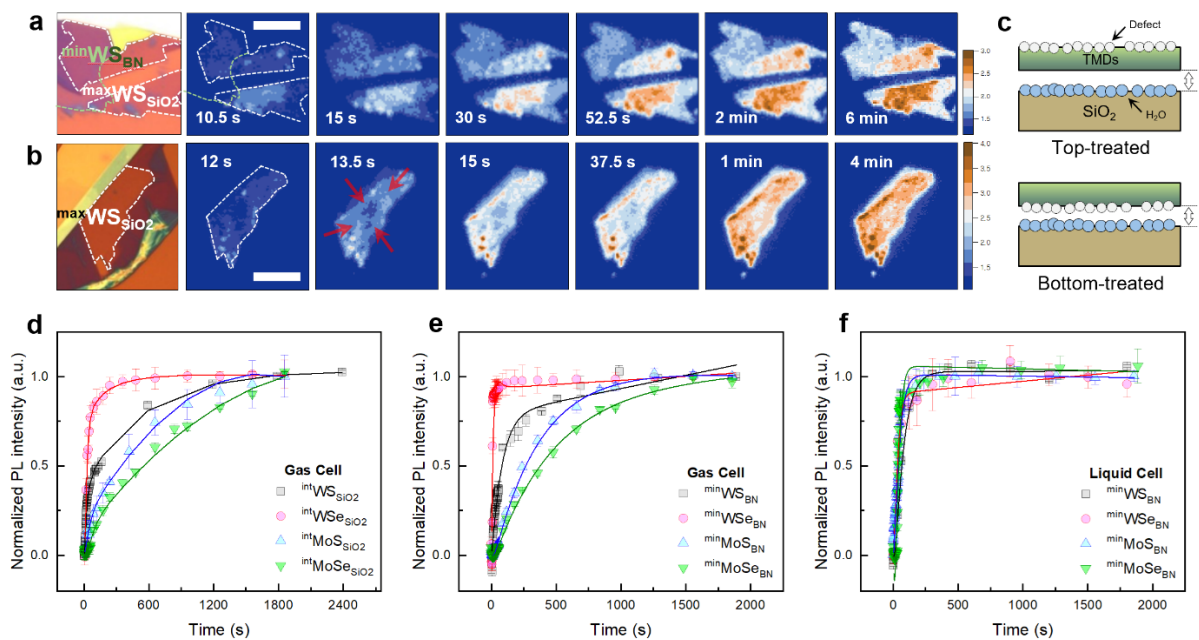

**Figure 4. CT reaction promoted by basal-plane defects.** (a & b) Optical micrographs (first column) and time-lapse wide-field PL enhancement images: $^{max}WS_{SiO2}$ and $^{min}WS_{BN}$ (a), and $^{max}WS_{SiO2}$ (b). The top (a) and bottom (b) surfaces of $WS_2$ were UVO-treated for 80 and 10 s, respectively. The red arrows in the third image of (b) denote the direction of changes. The samples were pre-equilibrated with Ar gas for 2 h before exposure to Ar:$O_2$ mixed gas at time zero. Scale bars are 7 and 10 μm, respectively. (c) Schematic side views of $^{max}WS_{SiO2}$ (from top to bottom) relevant for (a) and (b), respectively. (d ~ f) Normalized PL enhancement given as a function of $O_2$-exposure time: $^{int}MX_{SiO2}$ in the gaseous environment (d), $^{min}MX_{BN}$ in the gaseous environment (e), and $^{min}MX_{BN}$ in distilled water (f). Data were selected at each TMD area 2 ~ 3 μm off from edges. Error bars denote standard deviation obtained from several spots in raw PL images. Solid lines are guide to the eye.

***Role of basal-plane defects in CT.*** We found that structural defects in TMDs serve as reaction centers for CT. In Fig. 4a and 4b, we present how defects affect PL enhancement images of one sample that contained $^{min}WS_{BN}$ and $^{max}WS_{SiO2}$ areas. To create oxide-based defects on the front basal planes, we treated the sample with UV-generated ozone (UVO).[49] The treatment time ($t_{UVO}$) was chosen for the maximum PL intensity in the ambient conditions. As shown in Fig. S5, the integrated PL intensity obtained in the air reached a maximum at $t_{UVO}$ = 60 s. Despite the prominent defect-induced PL peak (Fig. S6)[50] for extended treatments, Raman spectra remained intact (Fig. S7), indicating that the density of defects is still very low.[51-53] Upon exposure to $O_2$, the PL enhancement of $^{min}WS_{BN}$ rapidly reached ~1.5 at 30 s (Fig. 4a), which stood in stark contrast to the UVO-untreated cases (Fig. 3b). Whereas a similar promotion by defects was observed for $^{max}WS_{SiO2}$, the edge-to-center progression could not be seen (Fig. 4a). These facts suggest that the change in Fig. 4a is dominated by CT occurring on the front basal plane containing UVO-generated defects. To corroborate this, we prepared $^{max}WS_{SiO2}$ samples with its UVO-treated $WS_2$ surface facing the substrates as schematically shown in Fig. 4c (see Methods & Fig. S8). Remarkably, the sample in Fig. 4b exhibited a similar degree of promotion but with evident directional propagation. The rate of spatial propagation for $^{max}WS_{SiO2}$ was even faster than without UVO treatments (Fig. 3b & 4b). These results led us to conclude that CT is more efficient with increasing density of defects, the role of which will be explained below.

We also revealed that the apparent CT kinetics varies drastically among materials, and the variation is essentially governed by the degree of hydration at reaction centers. Figure 4d presents the temporal change in the PL enhancement kinetics of four types of $^{int}MX_{SiO2}$, where M and X are Mo or W, and S or Se, respectively. The enhancement rose the fastest (slowest) for $WSe_2$ ($MoSe_2$) with the two sulfides in the middle. To exclude the role of interfacial diffusion, we also tested four $^{min}MX_{BN}$ samples. As shown in Fig. 4e, the material-dependence was maintained, which indicated that the difference originates from their material properties, not interfacial geometry. In Fig. 4f, we show $O_2$-induced PL enhancement of the four $^{min}MX_{BN}$ samples immersed in liquid water. Surprisingly, all the data collapsed onto an almost single fast varying curve. This result directly proves that the CT reaction in Figs. 4e & 4f was limited by the deficiency of hydrating water molecules.

There are a few factors that may play a role in the apparent material-dependence of the gas-phase CT (Fig. 4d). Most of all, the local hydrophilicity of basal planes may vary among the four TMDs because of their differential defect density and consequently affect their CT rates. This is strongly supported by the material-independent CT kinetics in liquid water (Fig. 4f). Whereas defect-free TMD crystals are expected to be hydrophobic,[54] the hydrophilicity was higher for 1L TMDs with more growth-related defects.[55] We also note that the difference in the water contact angle between 1L $WS_2$ and $MoS_2$ was slight but varied among studies,[54] which implies significant variance in the density of native defects. Moreover, the UVO-generated defects are expected to increase the degree of hydration at reaction centers and therefore the CT rate, because they are essentially metal oxides with polar character.[56] Second, defects may serve as CT centers, and their efficiency may vary among the four TMDs. Indeed, $MoS_2$ doped with hetero atoms was demonstrated as enhanced ORR catalysts.[57] As the CT-induced change in $\textbf{\textit{n}}_e$ is ~1x10$^{13}$ cm$^{-2}$ (Fig. 2a), the CT centers are likely to be single-atom vacancies, which are the most abundant (~1x10$^{13}$ cm$^{-2}$) defects in TMDs.[58] As the basal-plane defects accelerate CT (Fig. 4a), the kinetic variation in Fig. 4d suggests that the four TMDs are distinctive in the nature and number density of their native defects.

Third, CT within the Marcus-Gerischer model leads to material-dependence because the electron donor levels of the four TMDs are unequal.[59]

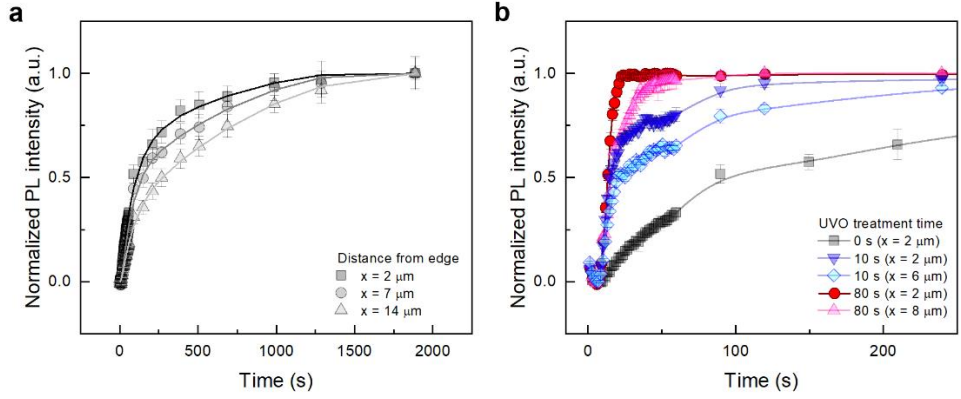

**Figure 5. Diffusional kinetics and defect-enhanced CT.** (a & b) Normalized PL enhancement given as a function of $O_2$-exposure time: pristine $^{max}WS_{SiO2}$ (a) and $^{max}WS_{SiO2}$ with artificial defects (b). The data in (a) were obtained from the three spots (P1 ~ P3) in Fig. 3b. The $WS_2$ surfaces facing the substrate were UVO-treated to create additional defects in (b). One data set of (a) was included in (b) for comparison. Error bars denote standard deviation. Solid lines are guide to the eye.

***Entangled kinetics of diffusion and CT.*** In Fig. 5a & 5b, we show the PL enhancement obtained from multiple positions of $^{max}WS_{SiO2}$ samples, prepared without and with UVO treatments of the bottom surface of $WS_2$. Most of all, the $O_2$-induced change was faster for the positions closer to edges (Fig. 5a), which is consistent with the edge-to-center changes observed in the wide-field images. When fitted with a simple exponential function, the rising time increased from 150 s for a near-edge position (x = 2 μm) to 340 s for a central one (x = 14 μm). Besides, the UVO treatments expedited the change (Fig. 5b), as also described in Figs. 3 & 4. The time constant for $^{max}WS_{SiO2}$ with UVO-generated defects was one order of magnitude smaller than that of pristine samples. Considering the finite gas exchange time (~15 s) of the optical gas cell, the kinetic difference between the pristine and UVO-treated samples will be even larger. The areas closer to edges exhibited faster rising for the UVO-treated samples as well.

We note that the spatiotemporal changes in the wide-field PL images are self-limited and result from the convolution between the interfacial diffusion and CT reaction. Whereas the two processes defy facile unequivocal isolation because of the entangled nature, the whole process can be described using the schemes given in Fig. 1d and 2c. As the first step, the interfacial diffusion of $O_2$ molecules is initiated when the molecules enter the interfacial space at the edges of TMD. In the limit that the length of a given edge is sufficiently larger than the diffusion length, the net molecular motion can be reduced to 1D diffusion along the direction normal to the edge and readily described by Fick's laws of diffusion. The CT rate at a given position is proportional to the local $O_2$ concentration,[25] but will eventually decrease to zero because of the lowering of TMD's Fermi level as the reaction proceeds.[10, 25] This explains the origin of the self-limited CT kinetics that $\mathbf{n_e}$ of TMD eventually converges to a constant. It is also notable that the CT rate is greatly enhanced (or reduction of the time constant) upon introduction of defects. As explained earlier, defects can contribute to the overall reaction by increasing the hydration level because of their polar nature and directly serving as a reaction center for ORR. Despite the entangled nature of the two processes, however, it is evident that the overall kinetics could be modulated by either geometric or chemical modification of the 2D space.

## Conclusions

In this work, we reported real-time observation of $O_2$ diffusion and ensuing CT involving oxygen reduction reaction occurring in the 2D space between 1L TMDs and dielectric substrates. As an optical probe sensitive to a trace amount of $O_2$, PL signals from TMDs were obtained in a time-lapse wide-field imaging mode. The overall rate of the two sequentially coupled processes could be enhanced by increasing the interfacial gap size and introducing artificial defects, respectively. We also showed that widely varying reaction kinetics of four TMDs was rate-determined by hydration required for the CT reaction. The in-situ wide-field PL images and extracted kinetic information provided the unprecedented mechanistic details of the two intimately coupled molecular processes in 2D space, which will shed light on designing efficient functional nanostructures and devices.

## ASSOCIATED CONTENT

## Supporting information

Change of relative humidity in the optical cell; Raw PL images with optical micrographs; AFM topographic images of $^{min}WS_{sapp}$; Diffusion-driven CT in $MoS_2$, $WSe_2$, and $MoSe_2$; Effects of UVO treatments on PL spectra of $^{int}WS_{SiO2}$; Defect-induced PL peak of UVO-treated $^{int}WS_{SiO2}$; Effects of UVO treatments on Raman spectra of $^{int}WS_{SiO2}$; Schematic procedure to prepare $^{max}MX_{SiO2}$ samples with either TMD surface UVO-treated; Spatial resolution of wide-field images.

## AUTHOR INFORMATION

**Corresponding author**

*E-mail: sunryu@postech.ac.kr

**Author contributions:** H.K. and S.R. designed the experiments. H.K. performed the experiments and data analysis. H.K. and S.R. wrote the manuscript.

**Competing interests:** We declare no competing interests.

## ACKNOWLEDGEMENT

S.R. acknowledges the financial support from Samsung Research Funding Center of Samsung Electronics under Project Number SSTF-BA1702-08. The authors thank Dr. Hyunjeong Kwon and Wontaek Kim for productive discussion on numerical analysis.

**TOC Graphics**

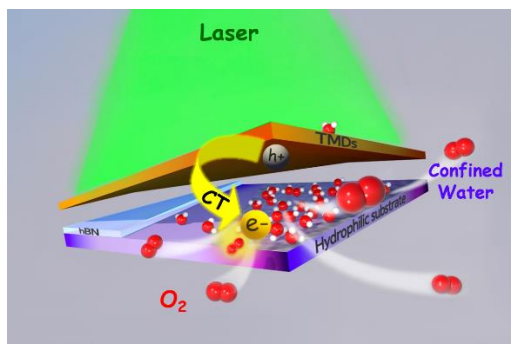